\documentclass[12pt]{article}
\usepackage{epsf}
\title{ {\bf
The radiative lepton flavor violating decays in the split fermion
scenario in the two  Higgs doublet model.}}
\author{\vspace{1cm}\\
        {\bf E. O. Iltan}
        \thanks{E-mail address:
        eiltan@newton.physics.metu.edu.tr}
 \\
        Physics Department, Middle East Technical University \\
        Ankara, Turkey\\}

\date{}

\begin{document}
\setlength{\baselineskip}{24pt}
\maketitle
\setlength{\baselineskip}{7mm}
\begin{abstract}
We study the branching ratios of the lepton flavor violating
processes $\mu\rightarrow e\gamma$, $\tau\rightarrow e\gamma$  and
$\tau\rightarrow \mu\gamma$ in the split fermion scenario, in the
framework of the two Higgs doublet model. We observe that the
branching ratios are relatively more sensitive to the
compactification scale and the Gaussian widths of the leptons in
the extra dimensions, for two extra dimensions and especially for
the $\tau\rightarrow \mu \gamma$ decay.
\end{abstract}
\thispagestyle{empty}
\newpage
\setcounter{page}{1}
\section{Introduction}
The radiative lepton flavor violating (LFV) decays exist at least
at one loop level and they are rich from the theoretical point of
view since the measurable quantities of these decays contain
number of free parameters of the model used. In framework of the
standard model (SM), these LFV decays are forbidden and, by
introducing the neutrino mixing with the non zero neutrino masses,
the LFV decays are allowed. However, their branching ratios (BRs)
are tiny and much below the experimental limits  due to the
smallness of neutrino masses. This forces one to go beyond the SM
and LFV decays are among the strong candidates for testing the
possible new physics effects.

There is an extensive experimental and theoretical work done on
the radiative LFV  decays in the literature. The current limits
for the branching ratios (BRs) of $\mu\rightarrow e\gamma$,
$\tau\rightarrow e\gamma$  and $\tau\rightarrow \mu\gamma$ decays
are $1.2\times 10^{-11}$ \cite{Brooks}, $3.9\times 10^{-7}$
\cite{Hayasaka} and $1.1\times 10^{-6}$ \cite{Ahmed},
respectively. The theoretical analysis of these decays has been
performed in various studies \cite{Barbieri1}-\cite{Chang}.
\cite{Barbieri1}-\cite{Barbieri7} were devoted to the analysis in
the supersymmetric models. In \cite{Iltan1, Iltan2, Diaz,
IltanExtrDim} and \cite{Chang}, they were examined in the
framework of the two Higgs doublet model (2HDM) and in a model
independent way respectively. With the extension of the Higgs
sector and the assumption that the flavor changing neutral
currents (FCNC's) at tree level are permitted, the BRs of the
radiative decays  under consideration can be enhanced
theoretically up to the experimental limits. 2HDM with tree level
FCNC currents is one of the candidate, and, in this model, the
radiative LFV decays are induced by the internal new neutral Higgs
bosons $h^0$ and $A^0$.

This work is devoted to the LFV processes $\mu\rightarrow
e\gamma$, $\tau\rightarrow e\gamma$  and $\tau\rightarrow
\mu\gamma$ in the 2HDM, with the inclusion of one (two) extra
spatial dimension. Here, we choose that the hierarchy of fermion
masses is coming from the overlap of the fermion Gaussian profiles
in the extra dimensions, so called the split fermion scenario
\cite{Hamed, Hamed2}. In this case the fermions are assumed to
locate at different points in the extra dimensions with the
exponentially small overlaps of their wavefunctions. There are
various works done on this scenario in the literature
\cite{Mirabelli}-\cite{Delgado10}. The explicit positions of left
and right handed components of fermions in the extra dimensions
have been predicted in \cite{Mirabelli}. Using the leptonic W
decays and the lepton violating processes, the restrictions on the
split fermions in the extra dimensions have been obtained in
\cite{Changg}. The CP violation in the quark sector has been
studied in \cite{Branco} and  to find stringent bounds on the size
of the compactification scale 1/R, the physics of kaon, neutron
and B/D mesons  has been analyzed in \cite{Chang2}. \cite{Hewett}
is devoted to the rare processes in the split fermion scenario and
\cite{Perez,Perez2} is related to the shapes and overlaps of the
fermion wave functions in the split fermion model. In
\cite{IltanEDMSplit} the electric dipole moments of charged
leptons have been predicted, respecting this scenario.

In the present work, we study the BRs of the radiative LFV decays
by considering that the leptons have Gaussian profiles in the
extra dimensions. First, we consider the BRs in the case of a
single extra dimension. In the following, we make the same
analysis when the number of extra dimensions is two and the
charged leptons are restricted to the fifth extra dimension, with
non-zero Gaussian profiles. Finally, we estimate the effects of
the extra dimensions if the non-zero Gaussian profiles exist in
both extra dimensions. We observe that the BRs are relatively more
sensitive to the compactification scale and the Gaussian widths of
the leptons in the extra dimensions, for two extra dimensions and
especially for the $\tau\rightarrow \mu \gamma$ decay.

The paper is organized as follows: In Section 2, we present the
BRs of the radiative LFV decays in the split fermion scenario, in
the 2HDM. Section 3 is devoted to discussion and our conclusions.
\section{The radiative lepton flavor violating decays in the split fermion
scenario in the two  Higgs doublet model }
The radiative LFV decays are worthwhile to study since they exist
at least in the loop level and they are rich theoretically. The
tiny  numerical values of the BRs of these decays in the SM forces
one to go beyond and the version of the 2HDM, permitting the
existence of the FCNCs and the LFV interactions at tree level, is
the simplest candidate. The new Yukawa couplings, which are
complex in general, play the main role in the calculation of the
physical quantities related to these decays. In addition to the
extension of the Higgs sector, the inclusion of the spatial extra
dimensions brings additional contributions. Here, we take the
effects of extra dimensions into account and we follow the idea
that the hierarchy of lepton masses are coming from the lepton
Gaussian profiles in the extra dimensions. The Yukawa Lagrangian
responsible for these interactions in a single extra dimension,
respecting the split fermion scenario, reads
\begin{eqnarray}
{\cal{L}}_{Y}=
\xi^{E}_{5 \,ij} \bar{\hat{l}}_{i L} \phi_{2} \hat{E}_{j R} + h.c.
\,\,\, , \label{lagrangian}
\end{eqnarray}
where $L$ and $R$ denote chiral projections $L(R)=1/2(1\mp
\gamma_5)$, $\phi_{2}$ is the new scalar doublet and $\xi^{E}_{5\,
ij}$ are the flavor violating complex Yukawa couplings in five
dimensions. Here, $\hat{l}_{i L}$ ($\hat{E}_{j R}$), with family
indices $i,j$, are the zero mode\footnote{Notice that we take only
the zero mode lepton fields in our calculations.} lepton doublets
(singlets) with Gaussian profiles in the extra dimension y and
they read
\begin{eqnarray}
\hat{l}_{i L}&=& N\,e^{-(y-y_{i L})^2/2 \sigma^2}\,l_{i L} ,
\nonumber
\\ \hat{E}_{j R}&=&N\, e^{-(y-y_{j R})^2/2 \sigma^2}\, E_{j R}\, ,
\label{gaussianprof}
\end{eqnarray}
with the normalization factor $N=\frac{1}{\pi^{1/4}\,
\sigma^{1/2}}$. $l_{i L}$ ($E_{j R}$) are the lepton doublets
(singlets) in four dimensions and the parameter $\sigma$
represents the Gaussian width of the leptons, satisfying the
property $\sigma << R$, where $R$ is the compactification radius.
In eq. (\ref{gaussianprof}), the parameters $y_{i L}$ and $y_{iR}$
are the fixed positions of the peaks of left and right handed
parts of $i^{th}$ lepton in the fifth dimension and they are
obtained by taking the observed lepton masses into account
\cite{Mirabelli}. The underlying idea is that the mass hierarchy
of leptons are coming from the relative positions of the Gaussian
peaks of the wave functions located in the extra dimension
\cite{Hamed,Hamed2,Mirabelli}. One possible set of locations for
the lepton fields in the fifth dimension read (see
\cite{Mirabelli} for details)
\begin{eqnarray}
P_{l_i}=\sqrt{2}\,\sigma\, \left(\begin{array}{c c c}
11.075\\1.0\\0.0
\end{array}\right)\,,\,\,\,\, P_{e_i}=\sqrt{2}\,\sigma\, \left(\begin{array}
{c c c} 5.9475\\4.9475\\-3.1498
\end{array}\right)
 \,\, . \label{location}
\end{eqnarray}

We choose the Higgs doublets $\phi_{1}$ and $\phi_{2}$ as
\begin{eqnarray}
\phi_{1}=\frac{1}{\sqrt{2}}\left[\left(\begin{array}{c c}
0\\v+H^{0}\end{array}\right)\; + \left(\begin{array}{c c} \sqrt{2}
\chi^{+}\\ i \chi^{0}\end{array}\right) \right]\, ;
\phi_{2}=\frac{1}{\sqrt{2}}\left(\begin{array}{c c} \sqrt{2}
H^{+}\\ H_1+i H_2 \end{array}\right) \,\, , \label{choice}
\end{eqnarray}
with the vacuum expectation values,
\begin{eqnarray}
<\phi_{1}>=\frac{1}{\sqrt{2}}\left(\begin{array}{c c}
0\\v\end{array}\right) \,  \, ; <\phi_{2}>=0 \,\, ,\label{choice2}
\end{eqnarray}
and collect SM (new) particles in the first (second) doublet.
Notice that $H_1$ and $H_2$ are the mass eigenstates $h^0$ and
$A^0$ respectively since no mixing occurs between two CP-even
neutral bosons $H^0$ and $h^0$ at tree level in our case. The LFV
interaction at tree level is carried by the the new Higgs field
$\phi_{2}$ and it is accessible to extra dimension. Following the
compactification on the orbifold $S^1/Z_2$, it reads
\begin{eqnarray}
\phi_{2}(x,y ) & = & {1 \over {\sqrt{2 \pi R}}} \left\{
\phi_{2}^{(0)}(x) + \sqrt{2} \sum_{n=1}^{\infty} \phi_{2}^{(n)}(x)
\cos(ny/R)\right\} \,, \label{SecHiggsField}
\end{eqnarray}
where $\phi_{2}^{(0)}(x)$ ($\phi_{2}^{(n)}(x)$) is  the Higgs
doublet in the four dimensions (the KK modes) including the
charged Higgs boson $H^+$ ($H^{(n)+}$), the neutral CP even-odd
Higgs bosons $h^0$- $A^0$ ($h^{0 (n)}$- $A^{0 (n)}$). The non-zero
$n^{th}$ KK mode of the charged Higgs mass is
$\sqrt{m_{H^\pm}^2+m_n^2}$, and the neutral CP even (odd) Higgs
mass is $\sqrt{m_{h^0}^2+m_n^2}$, ($\sqrt{m_{A^0}^2+m_n^2}$ ),
with the $n$'th level KK particle mass $m_n=n/R$.

\begin{figure}[htb]
\vskip 2.0truein \centering \epsfxsize=6.8in
\leavevmode\epsffile{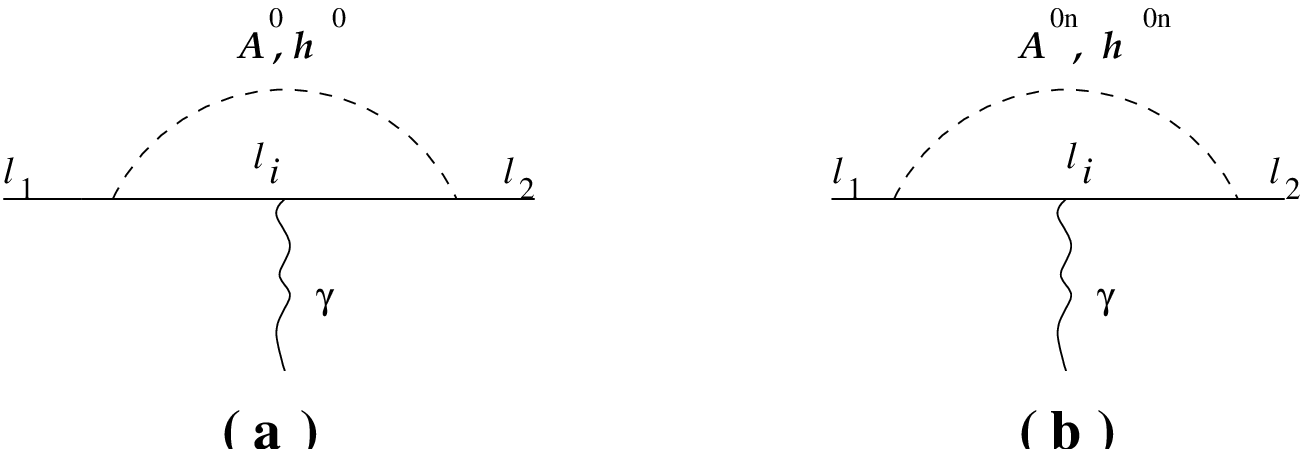} \vskip 1.0truein \caption[]{One loop
diagrams contribute to $l_1\rightarrow l_2 \gamma$ decay  due to
the zero mode (KK mode) neutral Higgs bosons $h^0$ and $A^0$
($h^{0 n}$ and $A^{0 n}$) in the 2HDM, for a single extra
dimension.} \label{fig1}
\end{figure}

The radiative decays under consideration exist at least in the one
loop level with the help of the intermediate neutral Higgs bosons
$h^0$, $A^0$ and their KK modes (see Fig. \ref{fig1}). The
lepton-lepton-$S$ vertex factors  $V^n_{LR\,(RL)\,ij}$ in the
vertices $\bar{\hat{f}}_{iL\, (R)}\,S^{(n)}(x)\,\cos(ny/R)\,
\hat{f}_{j R\, (L)}$, with $S=h^0,A^0$ and the right (left) handed
$i^{th}$ flavor lepton fields $\hat{f}_{j R\, (L)}$ in five
dimensions (see eq. (\ref{gaussianprof})), are obtained by the
integration over the fifth dimension and they read

%
\begin{eqnarray}
V^n_{LR\,(RL)\,ij}=e^{-n^2\,\sigma^2/4\,R^2}\,e^{-(y_{i L\,
(R)}-y_{i R\, (L)})^2/4 \sigma^2}\, \cos\, [\frac{n\,(y_{i L\,
(R)}+y_{i R\, (L)})}{2\,R}] \,\, . \label{Vij1}
\end{eqnarray}
In the case of $n=0$, the factor becomes
$V^0_{LR\,(RL)ij}=e^{-(y_{i L\, (R)}-y_{i R\, (L)})^2/4 \sigma^2}$
and we define the Yukawa couplings in four dimensions as
\begin{eqnarray}
\xi^{E}_{ij}\,\Big((\xi^{E \dagger}_{ij})^\dagger\Big)=
V^0_{LR\,(RL)\,ij} \, \xi^{E}_{5\, ij}\,\Big((\xi^{E}_{5\,
ij})^\dagger\Big)/\sqrt{2 \pi R} \,\, . \label{coupl4}
\end{eqnarray}

Since the LFV processes , $\mu\rightarrow e\gamma$,
$\tau\rightarrow e\gamma$ and $\tau\rightarrow \mu\gamma$ exist at
loop level, there appear the Logarithmic divergences in the
calculations and we eliminate them by using the on-shell
renormalization scheme
\footnote{Notice that, in this scheme, the self energy diagrams
for on-shell leptons vanish since they can be written as $
\sum(p)=(\hat{p}-m_{f_1})\bar{\sum}(p) (\hat{p}-m_{f_2})\, , $
however, the vertex diagrams $a$ and $b$ in Fig. \ref{fig1} give
non-zero contribution. In this case, the divergences can be
eliminated by introducing a counter term $V^{C}_{\mu}$ with the
relation $V^{Ren}_{\mu}=V^{0}_{\mu}+V^{C}_{\mu} \, , $ where
$V^{Ren}_{\mu}$ ($V^{0}_{\mu}$) is the renormalized (bare) vertex
and by using the gauge invariance: $k^{\mu} V^{Ren}_{\mu}=0$.
Here, $k^\mu$ is the four momentum vector of the outgoing
photon.}. Taking only $\tau$ lepton for the internal line
\footnote{We take into account only the internal $\tau$-lepton
contribution since, we respect the Sher scenerio \cite{Sher},
results in the couplings $\bar{\xi}^{E}_{N, ij}$ ($i,j=e,\mu$),
are small compared to $\bar{\xi}^{E}_{N,\tau\, i}$
$(i=e,\mu,\tau)$, due to the possible proportionality of them to
the masses of leptons under consideration in the vertices. Here,
we use the dimensionful coupling $\bar{\xi}^{E}_{N,ij}$ with the
definition $\xi^{E}_{N,ij}=\sqrt{\frac{4\, G_F}{\sqrt{2}}}\,
\bar{\xi}^{E}_{N,ij}$ where N denotes the word "neutral".}, the
decay width $\Gamma$, including a single extra dimension, reads as
\begin{eqnarray}
\Gamma (f_1\rightarrow f_2\gamma)=c_1(|A_1|^2+|A_2|^2)\,\, ,
\label{DWmuegam}
\end{eqnarray}
where
\begin{eqnarray}
A_1&=&Q_{\tau} \frac{1}{48\,m_{\tau}^2} \Bigg (6\,m_\tau\,
\bar{\xi}^{E *}_{N,\tau f_2}\, \bar{\xi}^{E *}_{N,f_1\tau}\, \Big(
F (z_{h^0})-F (z_{A^0})+ 2\,
\sum_{n=1}^{\infty}\,e^{-n^2\,\sigma^2/2\,R^2}\, c'_n
\,(f_1,\tau)\,c_n \,(f_2,\tau) \nonumber \\
\,\,\,\,\,\,\, &\times& \!\!\!(F (z_{n, h^0})-F (z_{n, A^0}))\Big
) + m_{f_1}\,\bar{\xi}^{E *}_{N,\tau f_2}\, \bar{\xi}^{E}_{N,\tau
f_1}\, \Big(G (z_{h^0})+G(z_{A^0})\nonumber \\
&+& 2\,\sum_{n=1}^{\infty}\,e^{-n^2\,\sigma^2/2\,R^2}\,c_n
\,(f_1,\tau)\,c_n \,(f_2,\tau)\, (G (z_{n, h^0})+G (z_{n,
A^0}))\Big) \Bigg)
\nonumber \,\, , \\
A_2&=&Q_{\tau} \frac{1}{48\,m_{\tau}^2} \Bigg (6\,m_\tau\,
\bar{\xi}^{E}_{N, f_2 \tau}\, \bar{\xi}^{E}_{N,\tau f_1}\, \Big(F
(z_{h^0})-F(z_{A^0})+2\,\sum_{n=1}^{\infty}\,e^{-n^2\,\sigma^2/2\,R^2}
\,c'_n \,(f_2,\tau)\,c_n \,(f_1,\tau)\nonumber \\
\,\,\,\,\,\,\, &\times& \!\!\! (F (z_{n, h^0})-F (z_{n, A^0}))\Big
)+ m_{f_1}\,\bar{\xi}^{E}_{N,f_2\tau}\, \bar{\xi}^{E *}_{N,f_1
\tau}\, \Big( G (z_{h^0})+G (z_{A^0}) \nonumber \\
&+& 2\,\sum_{n=1}^{\infty}\,e^{-n^2\,\sigma^2/2\,R^2} \,c'_n
\,(f_2,\tau)\,c'_n \,(f_1,\tau)\, (G (z_{n, h^0})+ G (z_{n, A^0}))
\Big) \Bigg)
 \,\, , \label{A1A2}
\end{eqnarray}
%
for $f_1\,(f_2)=\tau;\mu\,(\mu$ or $e; e)$. Here $c_1=\frac{G_F^2
\alpha_{em} m^3_{f_1}}{32 \pi^4}$, $A_1$ ($A_2$) is the left
(right) chiral amplitude, $z_{S}=\frac{m^2_{\tau}}{m^2_{S}}$,
$z_{n, S}=\frac{m^2_{\tau}}{m^2_{S}+(n/R)^2}$, $Q_{\tau}$ is the
charge of $\tau$ lepton and the parameters $c_n$, $c'_n$ read
\begin{eqnarray}
c_n \,(f,\tau)&=&\cos[\frac{n\,(y_{f R}+y_{\tau L})}{2\, R}]\,
\,, \nonumber \\
c'_n \,(f,\tau)&=&\cos[\frac{n\,(y_{f L}+y_{\tau R})}{2\, R}]\, .
\label{coeff}
\end{eqnarray}
In eq. (\ref{A1A2}) the functions $F_1 (w)$, $F_2 (w)$ are  given
by
\begin{eqnarray}
F (w)&=&\frac{w\,(3-4\,w+w^2+2\,ln\,w)}{(-1+w)^3} \, , \nonumber \\
G (w)&=&\frac{w\,(2+3\,w-6\,w^2+w^3+ 6\,w\,ln\,w)}{(-1+w)^4} \,\,
. \label{functions2}
\end{eqnarray}

Now, we present the amplitudes $A_1$ and $A_2$ appearing in the
decay width $\Gamma$ of the radiative LFV decays $f_1\rightarrow
f_2 \gamma$ (see eq. (\ref{DWmuegam})) in the case of two extra
dimensions (see Appendix section for details)
\begin{eqnarray}
A_1&=&Q_{\tau} \frac{1}{48\,m_{\tau}^2} \Bigg (6\,m_\tau\,
\bar{\xi}^{E *}_{N,\tau f_2}\, \bar{\xi}^{E *}_{N,f_1\tau}\, \Big(
F (z_{h^0})-F (z_{A^0})+4\,
\sum_{n,s}^{\infty}\,e^{-(n^2+s^2)\,\sigma^2/2\,R^2}\,
c'_{2\,(n,s)}\,(f_1,\tau) \nonumber \\
\,\,\,\,\,\,\, &\times& \!\!\!c_{2\,(n,s)} \,(f_2,\tau)\, (F
(z_{(n,s), h^0})-F (z_{(n,s), A^0}))\Big ) + m_{f_1}\,\bar{\xi}^{E
*}_{N,\tau f_2}\, \bar{\xi}^{E}_{N,\tau f_1}\, \Big(G (z_{h^0})+G
(z_{A^0}) \nonumber \\&+&
4\,\sum_{n,s}^{\infty}\,e^{-(n^2+s^2)\,\sigma^2/2\,R^2}\,c_{2\,(n,s)}
\,(f_1,\tau)\,c_{2\,(n,s)} \,(f_2,\tau)\, (G (z_{(n,s), h^0})+G
(z_{(n,s), A^0}))\Big) \Bigg)
\nonumber \,\, , \\
A_2&=&Q_{\tau} \frac{1}{48\,m_{\tau}^2} \Bigg (6\,m_\tau\,
\bar{\xi}^{E}_{N, f_2 \tau}\, \bar{\xi}^{E}_{N,\tau f_1}\, \Big(F
(z_{h^0})-F(z_{A^0})+4\,\sum_{n,s}^{\infty}\,e^{-(n^2+s^2)\,\sigma^2/2\,R^2}
\,c'_{2\,(n,s)} \,(f_2,\tau)\nonumber \\
\,\,\,\,\,\,\, &\times& \!\!\!c_{2\,(n,s)} \,(f_1,\tau)\, (F
(z_{(n,s), h^0})-F (z_{(n,s), A^0}))\Big )+
m_{f_1}\,\bar{\xi}^{E}_{N,f_2\tau}\, \bar{\xi}^{E *}_{N,f_1
\tau}\, \Big( G (z_{h^0})+G (z_{A^0})\nonumber \\&+&
4\,\sum_{n,s}^{\infty}\,e^{-(n^2+s^2)\,\sigma^2/2\,R^2}
\,c'_{2\,(n,s)} \,(f_2,\tau)\,c'_{2\,(n,s)} \,(f_1,\tau)\, (G
(z_{(n,s), h^0})+ G (z_{(n,s), A^0})) \Big) \Bigg)
 \,\, , \label{A1A22}
\end{eqnarray}
where the parameter $z_{(n,s), S}$ is defined as, $z_{(n,s),
S}=\frac{m^2_{\tau}}{m^2_{S}+n^2/R^2+s^2/R^2}$. In eq.
(\ref{A1A22}) the summation would be done over $n,s=0,1,2 ...$,
except $n=s=0$. Furthermore the parameters
$c_{2\,(n,s)}\,(f,\tau)$ and $c'_{2\,(n,s)}\,(f,\tau)$ read
\begin{eqnarray}
c_{2\,(n,s)}\, (f,\tau)&=&\cos[\frac{n\,(y_{f R}+y_{\tau L})+
s\,(z_{f R}+z_{\tau L})}{2\, R}]\, \,,
\nonumber \\
c'_{2\,(n,s)}(f,\tau)&=&\cos[\frac{n\,(y_{f L}+y_{\tau R}) +
s\,(z_{f L}+z_{\tau R})}{2\, R}]\, . \label{coeff22}
\end{eqnarray}
\section{Discussion}
The radiative LFV decays $f_1\rightarrow f_2\gamma$ exist at least
in the one loop level in the 2HDM where the tree level FCNC
interactions are permitted. The Yukawa couplings
$\bar{\xi}^E_{N,ij}, \, i,j=e, \mu, \tau$, which are the free
parameters of the model, play an essential role on the physical
parameters of these decays. Since we follow the idea that the
hierarchy of lepton masses are due to the lepton Gaussian profiles
in the extra dimensions, there appear exponential suppression
factors, originated from the different locations of various
flavors and their left and right handed parts of lepton fields, in
the Yukawa part of the lagrangian, after the integration of the
extra dimension(s) (see the eq. (\ref{Vij1}) (eq. (\ref{Vij2}))
for $n=0$ ($n,s=0$)). We take the Yukawa couplings in four
dimensions as the combination of these new factors and the higher
dimensional one (see eq. (\ref{coupl4}) and (\ref{coupl44})) and
consider that the couplings $\bar{\xi}^{E}_{N,ij},\, i,j=e,\mu $
are smaller compared to $\bar{\xi}^{E}_{N,\tau\, i}\,
i=e,\mu,\tau$ since latter ones contain heavy flavor. Furthermore,
we assume that, in four dimensions, the couplings
$\bar{\xi}^{E}_{N,ij}$ is symmetric with respect to the indices
$i$ and $j$.

Our analysis is devoted to the prediction of the effects of the
extra dimensions on the LFV radiative decays. Here we choose the
appropriate numerical values for the Yukawa couplings, by
respecting the current experimental measurements of these decays
(see Introduction section) and the muon anomalous magnetic moment
(see \cite{Iltananomuon} and references therein). Notice that, for
the Yukawa coupling $\bar{\xi}^{E}_{N,\tau \tau}$, we use the
numerical value which is greater than the upper limit of
$\bar{\xi}^{E}_{N,\tau \mu}$. We also study the effects of the
parameter $\rho=\sigma/R$, where $\sigma$ is the Gaussian width of
the fermions (see \cite{Mirabelli} for details). For the
compactification scale $1/R$, there are numerous constraints for a
single extra dimension in the split fermion scenario. The direct
limits from searching for KK gauge bosons imply $1/R> 800\,\,
GeV$. The precision electro weak bounds on higher dimensional
operators generated by KK exchange place a far more stringent
limit $1/R> 3.0\,\, TeV$ \cite{Rizzo}. In \cite{Hewett}, the lower
bounds for the scale $1/R$ have been obtained as $1/R > 1.0 \,\,
TeV$ from $B\rightarrow \phi \, K_S$, $1/R > 500\,\, GeV$ from
$B\rightarrow \psi \, K_S$ and $1/R > 800\,\, GeV$ from the upper
limit of the $BR$, $BR \, (B_s \rightarrow \mu^+ \mu^-)<
2.6\,\times 10^{-6}$. We make our analysis by choosing an
appropriate range for the compactification scale $1/R$, by
respecting these limits in the case of a single extra dimension.
For two extra dimensions we used the same broad range for $1/R$.

In the model we use there are various free parameters, the new
Yukawa couplings, the masses of new Higgs bosons and the ones
coming from the split fermion scenario, namely, the
compactification scale and the possible locations of fermions in
the extra dimensions. It is obvious that the predictions are
sensitive to those parameters and it is not easy to decide whether
the enhancement comes from the broad region of one parameter set
belonging to the 2HDM part, or the one belonging to the split
fermion scenario. However, with the possible forthcoming
experimental results of the processes which are more sensitive to
the new parameters coming from the new Higgs doublet, the more
stringent restrictions can be obtained. This would lead to more
accurate analysis of the effects due to the split fermion
scenario. In our case, we try to estimate the sensitivity of the
BRs to the split fermion scenario by fixing the other parameters
and we expect that the more accurate discussion of this scenario
can be reached with forthcoming  experiments which reduces the
sensitivities at several orders.

In the present work, we take split leptons in a single and two
extra dimensions and use a possible set of locations to calculate
the extra dimension contributions. We make the analysis in one and
two extra dimensions. In the case of a single extra dimension (two
extra dimensions) we use the estimated location of the leptons
given in eq. (\ref{location}) (eq. (\ref{location2})) to calculate
the lepton-lepton-Higgs scalar KK mode vertices. For two extra
dimensions, first, we take that the leptons are restricted to the
fifth extra dimension, with non-zero Gaussian profiles. This is
the case that the enhancement in the BRs of the present decays is
relatively large. The reason beyond the enhancement is the well
known KK mode abundance of Higgs fields. Finally, we assume that
the leptons have non-zero Gaussian profiles also in the sixth
dimension and using a possible set of locations in the fifth and
sixth extra dimensions (see eq. (\ref{location2})), we calculated
BRs of the decays under consideration. In this case the additional
the exponential factor appearing in the second summation further
suppresses the BRs.

\begin{figure}[htb]
\vskip -3.0truein \centering \epsfxsize=6.8in
\leavevmode\epsffile{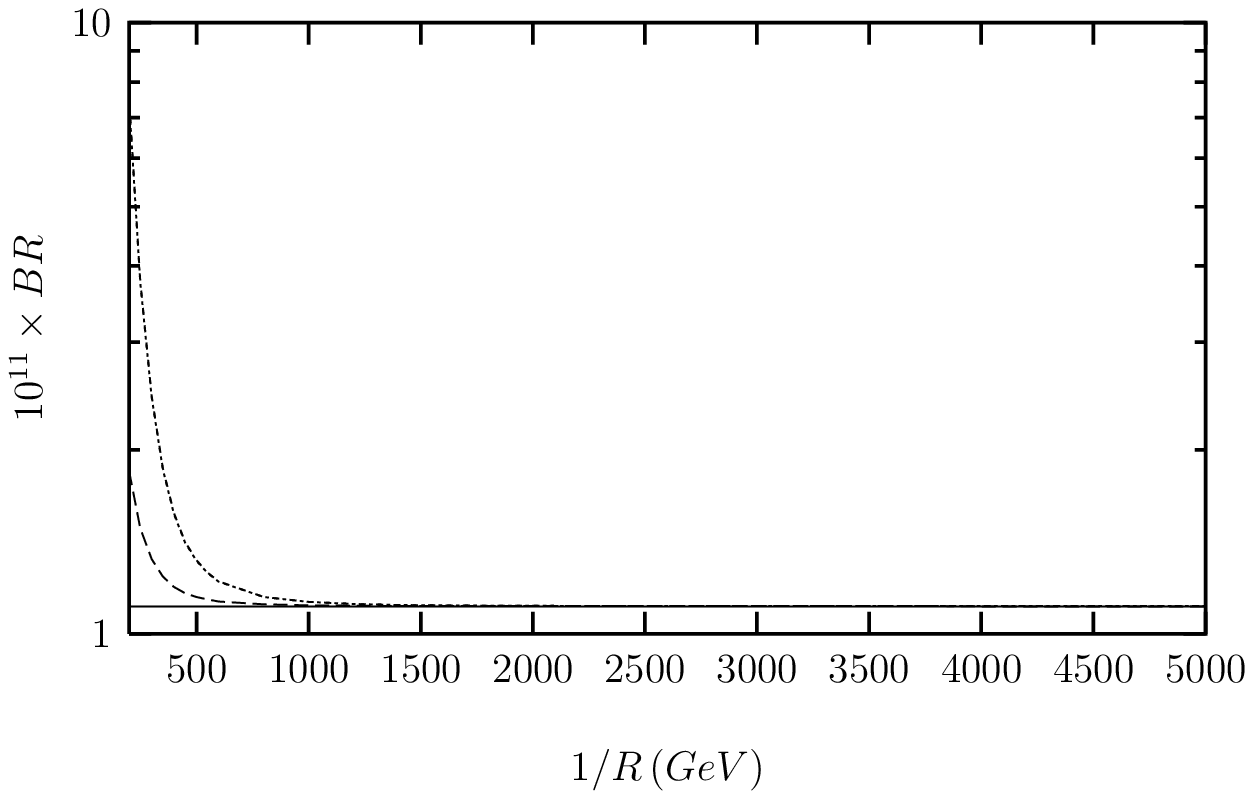} \vskip -3.0truein
\caption[]{$BR(\mu\rightarrow e \gamma$) with respect to the scale
$1/R$ for $\rho=0.01$, $m_{h^0}=100\, GeV$, $m_{A^0}=200\, GeV$
and the real couplings $\bar{\xi}^{E}_{N,\tau \mu} =10\, GeV$,
$\bar{\xi}^{E}_{N,\tau e} =0.001\, GeV$. Here the solid (dashed,
small dashed, dotted) line represents the BR without extra
dimension (with a single extra dimension, with two extra
dimensions where the leptons have non-zero Gaussian profiles in
the fifth extra dimension, with two extra dimensions where the
leptons have non-zero Gaussian profiles in both extra
dimensions).} \label{BRmuegamR}
\end{figure}

\begin{figure}[htb]
\vskip -3.0truein \centering \epsfxsize=6.8in
\leavevmode\epsffile{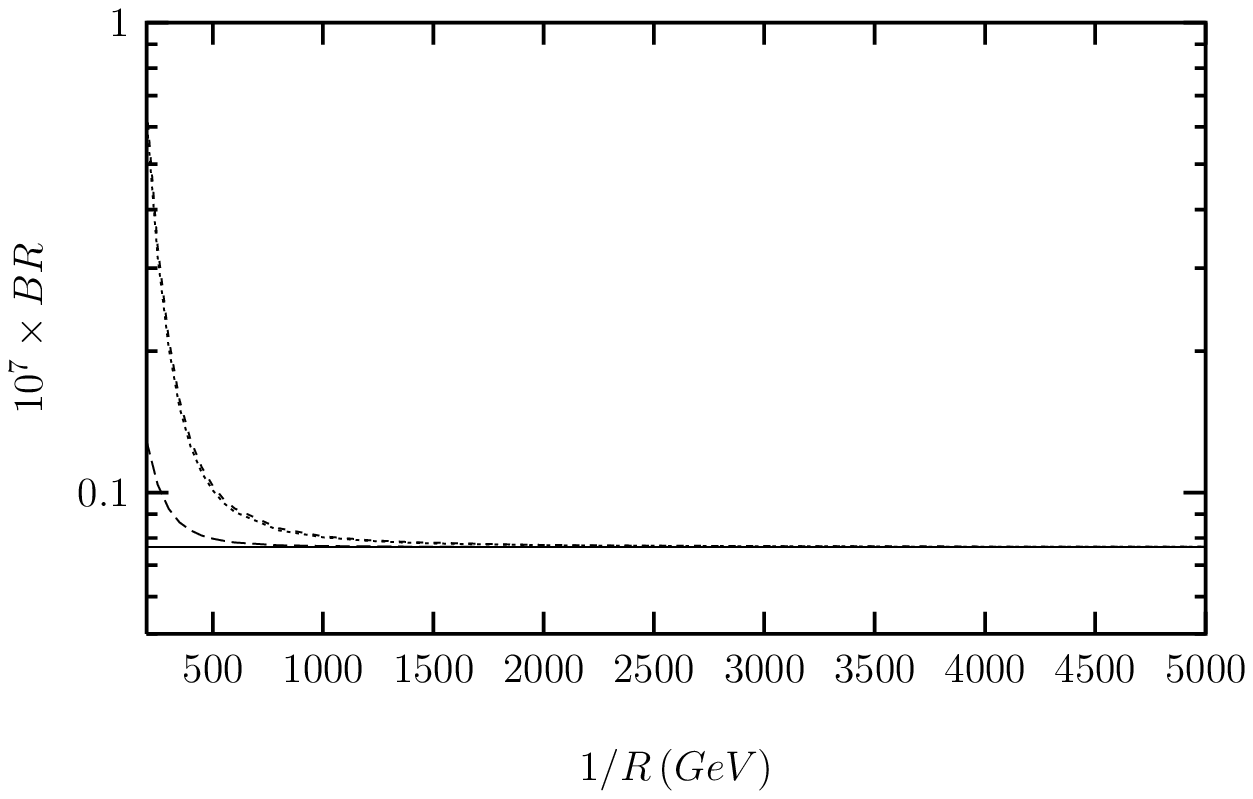} \vskip -3.0truein
\caption[]{BR($\tau\rightarrow e \gamma$) with respect to the
scale $1/R$ for $\rho=0.01$, $m_{h^0}=100\, GeV$, $m_{A^0}=200\,
GeV$ and the real couplings $\bar{\xi}^{E}_{N,\tau \tau} =100\,
GeV$, $\bar{\xi}^{E}_{N,\tau e} =1\, GeV$. Here the solid (dashed,
small dashed, dotted) line represents the BR without extra
dimension (with a single extra dimension, with two extra
dimensions where the leptons have non-zero Gaussian profiles in
the fifth extra dimension, with two extra dimensions where the
leptons have non-zero Gaussian profiles in both extra
dimensions).} \label{BRtauegamR}
\end{figure}

\begin{figure}[htb]
\vskip -3.0truein \centering \epsfxsize=6.8in
\leavevmode\epsffile{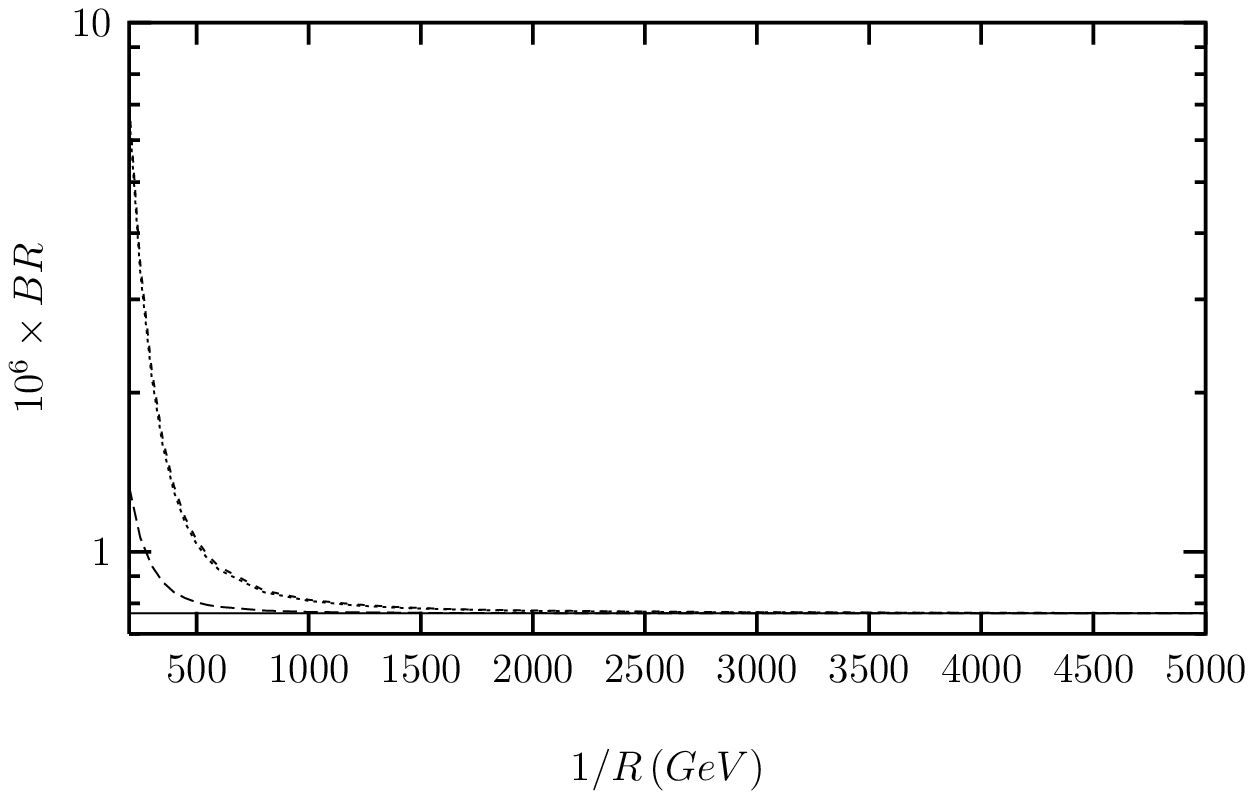} \vskip -3.0truein
\caption[]{BR($\tau\rightarrow \mu \gamma$) with respect to the
scale $1/R$ for $\rho=0.01$, $m_{h^0}=100\, GeV$, $m_{A^0}=200\,
GeV$ and the real couplings $\bar{\xi}^{E}_{N,\tau \tau} =100\,
GeV$, $\bar{\xi}^{E}_{N,\tau \mu} =10\, GeV$. Here the solid
(dashed, small dashed, dotted) line represents the BR without
extra dimension (with a single extra dimension, with two extra
dimensions where the leptons have non-zero Gaussian profiles in
the fifth extra dimension, with two extra dimensions where the
leptons have non-zero Gaussian profiles in both extra
dimensions).} \label{BRtaumugamR}
\end{figure}

In  Fig. \ref{BRmuegamR} (\ref{BRtauegamR} \textbf{;}
\ref{BRtaumugamR}), we plot the BR of the decay $\mu\rightarrow e
\gamma$ ($\tau\rightarrow e \gamma$ \textbf{;} $\tau\rightarrow
\mu \gamma$) with respect to the compactification scale $1/R$ for
$\rho=0.01$, $m_{h^0}=100\, GeV$, $m_{A^0}=200\, GeV$ and the real
couplings $\bar{\xi}^{E}_{N,\tau \mu} =10\, GeV$,
$\bar{\xi}^{E}_{N,\tau e} =0.001\, GeV$ ($\bar{\xi}^{E}_{N,\tau
\tau} =100\, GeV$, $\bar{\xi}^{E}_{N,\tau e} =1\, GeV$ \textbf{;}
$\bar{\xi}^{E}_{N,\tau \tau} =100\, GeV$, $\bar{\xi}^{E}_{N,\tau
\mu} =10\, GeV$)\footnote{For $\tau\rightarrow e \gamma$ we take
the numerical value of the coupling $\bar{\xi}^{E}_{N,\tau e}$,
$\bar{\xi}^{E}_{N,\tau e} =1\, GeV$. Here we try to reach the new
experimental result of the BR of this decay (see \cite{Hayasaka}).
With the more sensitive future measurements of the BRs of these
decays these couplings would be fixed more accurately.}. Here the
solid (dashed, small dashed, dotted) line represents the BR
without extra dimension (with a single extra dimension, with two
extra dimensions where the leptons have non-zero Gaussian profiles
in the fifth extra dimension, with two extra dimensions where the
leptons have non-zero Gaussian profiles in both extra dimensions).
It is observed that BR is weakly sensitive to the parameter $1/R$
for the $1/R>500\, GeV$ for a single extra dimension. The
enhancement of the BR is relatively larger for two extra
dimensions due to the Higgs scalar KK mode abundances and, in this
case, the sensitivity is weak for $1/R>1.0\, TeV$. However, these
contributions do not increase extremely due to the suppression
exponential factor appearing in the summations. Furthermore, the
numerical values of BRs are slightly greater in the case that the
leptons have non-zero Gaussian profiles only in the fifth extra
dimension.

Now, we would like to present the amount of the enhancements in
the BRs of the decays we study, by taking the existing bounds of
the compactification scale $1/R$ into account and to discuss the
possibility of observations of these additional contributions. For
$\mu\rightarrow e \gamma$ ($\tau\rightarrow e \gamma$ \textbf{;}
$\tau\rightarrow \mu \gamma$) decay, the enhancement in the BR for
a single extra dimension is at the order of $0.7\%$ ($1.0\%$
\textbf{;} $1.3\%$) for $1/R\sim 800\, GeV$, compared to the case
where there is no extra dimension. For the greater value of the
scale $1/R$, $1/R\sim 3.0\, TeV$, the enhancement reads $0.01\%$
($0.03\%$ \textbf{;} $0.05\%$). These numbers show that the
enhancement in the BR of the decay $\tau\rightarrow \mu \gamma$ is
larger compared to the BRs of the others and this decay is
probably the better candidate among the present LFV decays  to
detect the effects of the extra dimensions for a single extra
dimension. Notice that, for $1/R\sim 3.0\, TeV$, the enhancements
in the BRs of these decays are weak and difficult to observe.

For two extra dimensions, where the leptons have non-zero Gaussian
profiles in the fifth extra dimension, the enhancement in the BR
of the decay $\mu\rightarrow e \gamma$ ($\tau\rightarrow e \gamma$
\textbf{;} $\tau\rightarrow \mu \gamma$) is at the order of the
magnitude of $3.6\%$ ($9.5\%$ \textbf{;} $10.6\%$) for $1/R\sim
800\, GeV$, compared to the case that there is no extra dimension.
For $1/R\sim 3.0\, TeV$, the enhancement reads $0.05\%$ ($0.5\%$
\textbf{;} $0.6\%$) \footnote{For two extra dimensions where the
leptons have non-zero Gaussian profiles in both extra dimensions,
the enhancement in the BR is at the order of the magnitude of
$3.5\%$ ($8.6\%$ \textbf{;} $9.7\%$) for $1/R\sim 800\, GeV$, for
$\mu\rightarrow e \gamma$ ($\tau\rightarrow e \gamma$ \textbf{;}
$\tau\rightarrow \mu \gamma$) decay, compared to the case that
there is no extra dimension. For $1/R\sim 3.0\, TeV$, the
enhancement reads $0.05\%$ ($0.4\%$ \textbf{;} $0.5\%$).}. This
shows that the additional contributions cause to increase the BR
at the order of $\sim 10\%$ for $\tau\rightarrow e \gamma$ and
$\tau\rightarrow \mu \gamma$ decays, in the case of $1/R\sim 800\,
GeV$. Therefore, for two extra dimensions, $\tau\rightarrow e
\gamma$ and $\tau\rightarrow \mu \gamma$ decays can ensure
valuable information on the effects of extra dimensions in the
split fermion scenario. Notice that, the lower bound of the
compactification scale probably will be different in the case of
two extra dimensions, compared to the one existing for a single
extra dimension. However, in our calculations, we used the same
broad range of the scale $1/R$ for a single and two extra
dimensions.

\begin{figure}[htb]
\vskip -3.0truein \centering \epsfxsize=6.8in
\leavevmode\epsffile{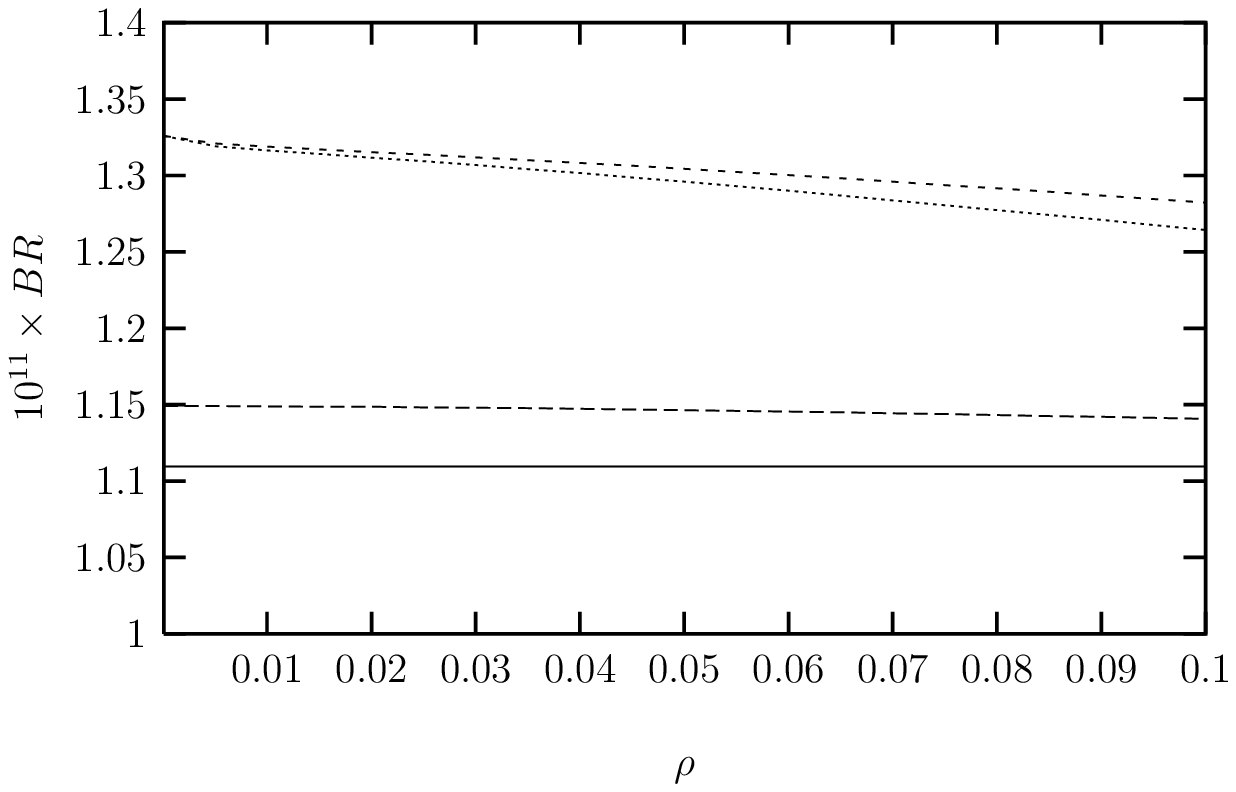} \vskip -3.0truein
\caption[]{The same as Fig. \ref{BRmuegamR} but with respect to
parameter $\rho$ and for $1/R=500\, GeV$.} \label{BRmuegamro}
\end{figure}

\begin{figure}[htb]
\vskip -3.0truein \centering \epsfxsize=6.8in
\leavevmode\epsffile{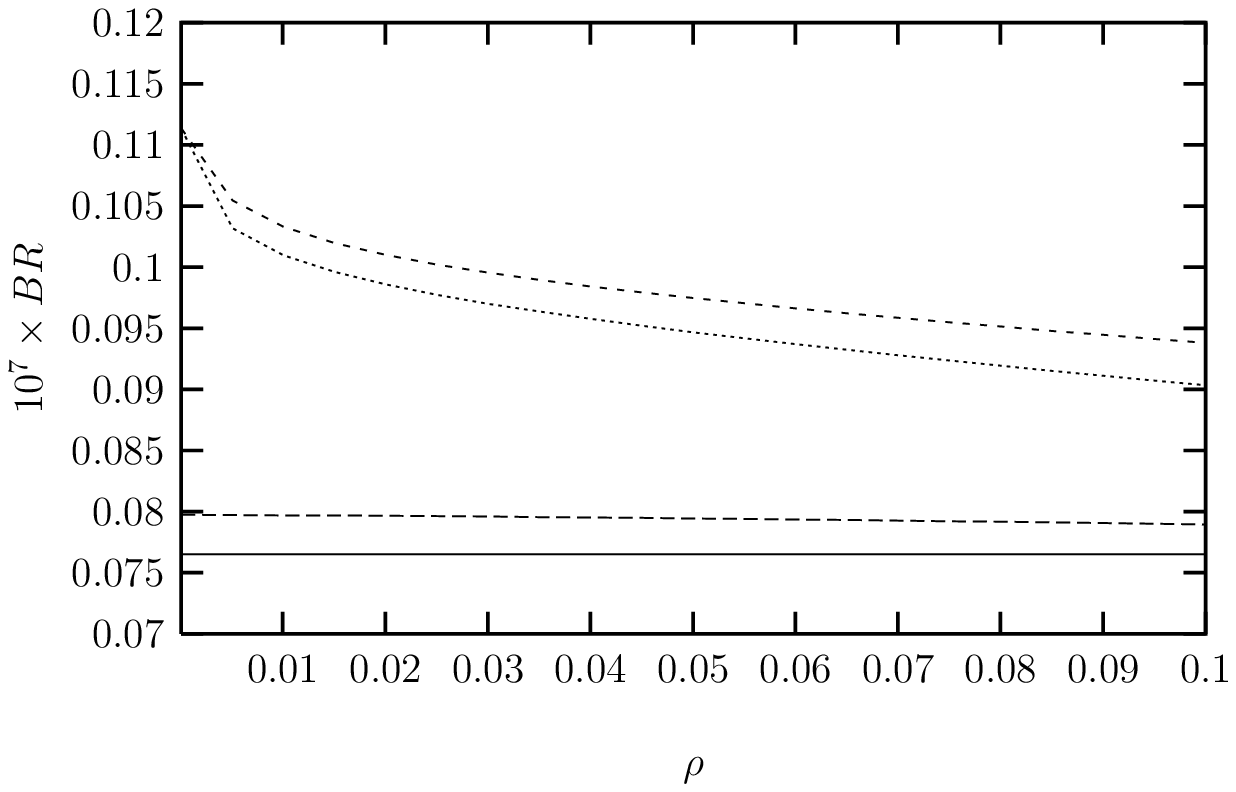} \vskip -3.0truein
\caption[]{The same as Fig. \ref{BRtauegamR} but with respect to
parameter $\rho$ and for $1/R=500\, GeV$.} \label{BRtauegamro}
\end{figure}

\begin{figure}[htb]
\vskip -3.0truein \centering \epsfxsize=6.8in
\leavevmode\epsffile{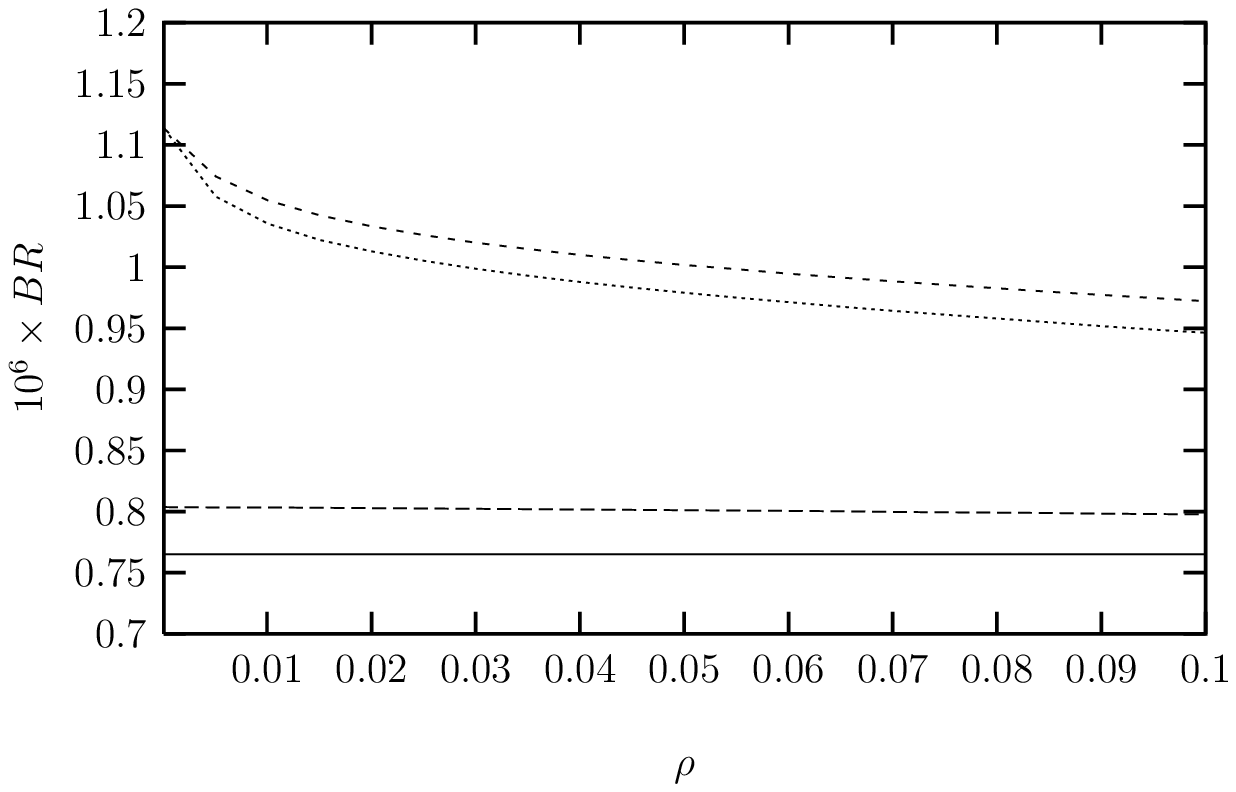} \vskip -3.0truein
\caption[]{The same as Fig. \ref{BRtaumugamR} but with respect to
parameter $\rho$ and for $1/R=500\, GeV$.} \label{BRtaumugamro}
\end{figure}

Fig. \ref{BRmuegamro} (\ref{BRtauegamro} \textbf{;}
\ref{BRtaumugamro}) is devoted to the parameter $\rho$ dependence
of the BR of the decay $\mu\rightarrow e \gamma$ ($\tau\rightarrow
e \gamma$ \textbf{;} $\tau\rightarrow \mu \gamma$) for $1/R=500\,
GeV$, $m_{h^0}=100\, GeV$, $m_{A^0}=200\, GeV$ and the real
couplings $\bar{\xi}^{E}_{N,\tau \mu} =10\, GeV$,
$\bar{\xi}^{E}_{N,\tau e} =0.001\, GeV$ ($\bar{\xi}^{E}_{N,\tau
\tau} =100\, GeV$, $\bar{\xi}^{E}_{N,\tau e} =1\, GeV$ \textbf{;}
$\bar{\xi}^{E}_{N,\tau \tau} =100\, GeV$, $\bar{\xi}^{E}_{N,\tau
\mu} =10\, GeV$). Here the solid (dashed, small dashed, dotted)
line represents the BR without extra dimension (with a single
extra dimension, with two extra dimensions where the leptons have
non-zero Gaussian profiles in the fifth extra dimension, with two
extra dimensions where the leptons have non-zero Gaussian profiles
in both extra dimensions). The BR of the decay $\mu\rightarrow e
\gamma$ ($\tau\rightarrow e \gamma$ \textbf{;} $\tau\rightarrow
\mu \gamma$) changes $0.8\%$ ($1.0\%$ \textbf{;} $0.8\%$) in the
given interval of the parameter $\rho$ and for $1/R=500\, GeV$, in
the case of a single extra dimension. This shows that the BRs of
the present decays are weakly sensitive to $\rho$, which is the
main parameter controlling the Gaussian widths and the possible
locations of the fermions in the extra dimensions. For two extra
dimensions, the sensitivity of the BRs to the parameter $\rho$
increases. For the decay $\mu\rightarrow e \gamma$
($\tau\rightarrow e \gamma$ \textbf{;} $\tau\rightarrow \mu
\gamma$) the BR changes $3.6\%$ ($19.0\%$ \textbf{;} $15.0\%$) in
the given interval of the parameter $\rho$ and for $1/R=500\,
GeV$, in the case of two extra dimensions where the leptons have
non-zero Gaussian profiles in the fifth extra dimension. For two
extra dimensions where the leptons have non-zero Gaussian profiles
in both extra dimensions, the BR of the decay $\mu\rightarrow e
\gamma$ ($\tau\rightarrow e \gamma$ \textbf{;} $\tau\rightarrow
\mu \gamma$) changes $5.1\%$ ($24.0\%$ \textbf{;} $18.0\%$) in the
given interval of the parameter $\rho$. It is shown that the
sensitivities of the BRs of studied LFV decays are relatively
greater for two extra dimensions and this sensitivity increases
for the case where the leptons have non-zero Gaussian profiles in
both extra dimensions. In addition to this, the BRs of tau decays
are sensitive to the parameter $\rho$ and this sensitivity
increases for $\rho<0.05$, in the case of two extra dimensions.

At this stage we would like to discus the possibilities of
detecting the additional contributions due to the extra dimensions
with the present and possible forthcoming experimental
measurements. The experimental work for the lepton flavor
violating decays has been done  since the discoveries of heavy
leptons. A new experiment, to search for the lepton flavor
violating decay $\mu\rightarrow e \gamma$ \cite{Nicolo}  at PSI
has been described and the aim of the experiment was to reach to a
sensitivity of $BR=10^{-14}$, improved by three order of
magnitudes with respect to previous searches. At present the
experiment (PSI-R-99-05 Experiment) to search the $\mu\rightarrow
e \gamma$ decay is still running in the MEG \cite{Yamada}. For the
$\tau\rightarrow \mu \gamma$ decay, recently, an upper limit of
$BR=9.0\, (6.8)\, 10^{-9}$ at $90\%$ CL has been obtained
\cite{Roney} (\cite{Aubert}) and this result is an improvement
almost by one order of magnitudes with respect to previous one.
The future measurement of the radiative $\mu\rightarrow e \gamma$
decay with the sensitivity of $BR=10^{-14}$, hopefully, would make
it possible to detect possible the additional contributions,
especially, coming from two extra dimensions, even for the scale
$1/R\sim 3.0\, TeV$. On the other hand, for the decay
$\tau\rightarrow \mu \gamma$, the enhancement in the BR in the
case of two extra dimensions is at the order of the magnitude of
$\sim 0.6\%$ for $1/R\sim 3.0\, TeV$, and the one order improved
experimental value of the BR would ensure a possible detection of
the extra dimension effects even for large values of the
compactification scale $1/R$.

As a summary, the BR is weakly sensitive to the parameter $1/R$
for $1/R>500\, GeV$ for a single extra dimension, however, this
sensitivity increases for two extra dimensions. The exponential
suppression factor appearing in the summations reduces the extreme
enhancement due to the Higgs KK mode abundance. Furthermore, the
BR is weakly sensitive to the parameter $\rho$ especially for a
single extra dimension case. In the two extra dimensions, this
sensitivity is slightly larger compared to the one in the single
extra dimension. With the help of the forthcoming most accurate
experimental measurements of the radiative LFV decays, especially
the $\tau\rightarrow \mu \gamma$ decay, the valuable information
can be obtained to detect the effects due to the extra dimensions
in the case of split fermion scenario.
\section{Acknowledgement}
This work has been supported by the Turkish Academy of Sciences in
the framework of the Young Scientist Award Program.
(EOI-TUBA-GEBIP/2001-1-8)
\newpage
\section{Appendix}
In the two extra dimensions, after the compactification on the
orbifold $(S^1\times S^1)/Z_2$, the new Higgs field $\phi_{2}$ is
expanded as
\begin{eqnarray}
\phi_{2}(x,y,z ) & = & {1 \over {2 \pi R}} \left\{
\phi_{2}^{(0,0)}(x) + 2 \sum_{n,s}^{\infty} \phi_{2}^{(n,s)}(x)
\cos(ny/R+sz/R)\right\} \,. \label{SecHiggsField2}
\end{eqnarray}
The KK modes of charged (neutral CP even, neutral CP odd) Higgs
fields  existing in the new Higgs doublet have the masses
$\sqrt{m_{H^\pm}^2+m_n^2+m_s^2}$, ($\sqrt{m_{h^0}^2+m_n^2+m_s^2}$,
\\ $\sqrt{m_{A^0}^2+m_n^2+m_s^2}$ ) where $m_n=n/R$ and $m_s=s/R$
are the masses of $n$'th and $s$'th level KK particles. If the
leptons are restricted only the fifth dimension, the vertex factor
$V^n_{LR\,(RL)\,ij}$ is the same as the one in eq. (\ref{Vij1}).
In the case that the leptons are accessible two both dimensions
with Gaussian profiles
\begin{eqnarray}
\hat{l}_{i L}&=& N\,e^{-\Big((y-y_{i L})^2+(z-z_{i L})^2\Big)/2
\sigma^2}\,l_{i L} , \nonumber
\\ \hat{E}_{j R}&=&N\, e^{-\Big((y-y_{j R})^2+(z-z_{j R})^2\Big)/2
\sigma^2}\, E_{j R}\, , \label{gaussianprof2}
\end{eqnarray}
and the normalization factor $N=\frac{1}{\pi^{1/2}\, \sigma}$,
the integration of the part of the Lagrangian \\
$\bar{\hat{f}}_{iL(R)}\,S^{(n,s)}(x)\,\cos(ny/R+sz/R)\, \hat{f}_{j
R (L)}$, with $S=h^0,A^0$, over the fifth and sixth extra
dimensions results in the vertex factor
\begin{eqnarray}
V^{(n,s)}_{LR\,(RL)\,ij}&=&e^{-(n^2+s^2)\,\sigma^2/4\,R^2}\,e^{-\Big(
(y_{i L\, (R)}-y_{j R\, (L)})^2+(z_{i L\, (R)}-z_{j R\, (L)})^2
\Big)/4 \sigma^2}\nonumber \\&\times& \cos\, [\frac{n\,(y_{i L\,
(R)}+y_{j R\, (L)})+s\,(z_{i L\, (R)}+z_{j R\, (L)})}{2\,R}] \,.
\label{Vij2}
\end{eqnarray}
Similar to a single extra dimension case, we define the Yukawa
couplings in four dimension as
\begin{eqnarray}
\xi^{E}_{ij}\,\Big((\xi^{E}_{ij})^\dagger\Big)=
V^{(0,0)}_{LR\,(RL)\,ij} \, \xi^{E}_{6 \,ij}\,\Big((\xi^{E}_{6
\,ij})^\dagger\Big)/2 \pi R\, , \label{coupl44}
\end{eqnarray}
where $V^{(0,0)}_{LR\,(RL)\,ij}=e^{-(y_{i L\, (R)}-y_{i R\,
(L)})^2-z_{i L\, (R)}-z_{i R\, (L)})^2/4 \sigma^2}$.

Now we present a possible positions of left handed and right
handed leptons in the two extra dimensions, by using the observed
masses \footnote{The calculation is similar to the one presented
in \cite{Mirabelli} which is done for a single extra dimension.}.
With the assumption that the lepton mass matrix is diagonal, one
of the possible set of locations for the Gaussian peaks of the
lepton fields in the two extra dimensions can be obtained as
\cite{IltanEDMSplit}
\begin{eqnarray}
P_{l_i}=\sqrt{2}\,\sigma\, \left(\begin{array}{c c c}
(8.417,8.417)\\(1.0,1.0)\\(0.0,0.0)
\end{array}\right)\,,\,\,\,\,
P_{e_i}=\sqrt{2}\,\sigma\, \left(\begin{array} {c c c}
(4.7913,4.7913)\\(3.7913,3.7913)\\(-2.2272,-2.2272)
\end{array}\right)
 \,\, . \label{location2}
\end{eqnarray}
where the numbers in the parenthesis denote the y and z
coordinates of the location of the Gaussian peaks of lepton
flavors in the extra dimensions. Here we choose the same numbers
for the y and z locations of the Gaussian peaks.

\newpage
\newpage
\end{document}